\documentstyle[12pt,aps]{revtex}
\newcommand{\bold}[1]{\mbox{\boldmath $#1$}}
\tighten
\begin{document}
\draft
\title{\flushleft Comment on `Vector potential of the Coulomb gauge'\footnote
{This comment is written by V Hnizdo in his private capacity. No support or
endorsement by the Centers for Disease Control and Prevention is intended or
should be inferred.
}}

\author{\flushleft V Hnizdo}

\address{\flushleft
National Institute for Occupational Safety and Health,\\
1095 Willowdale Road, Morgantown, WV 26505, USA
\newline
\newline
E-mail: vbh5@cdc.gov}

\address{\flushleft\rm
{\bf Abstract}. 
The expression for the Coulomb-gauge vector potential in terms
of the `instantaneous' magnetic field derived by Stewart
[2003 {\it Eur. J. Phys.} {\bf 24} 519] by employing Jefimenko's equation
for the magnetic field and Jackson's formula for the Coulomb-gauge
vector potential can be proven much more simply.}

\maketitle

\section*{}
\noindent                                                       
In a recent article \cite{Stew}, Stewart has derived the following expression
for the Coulomb-gauge vector potential $\bold{A}_{\rm C}$ in terms of the
`instantaneous' magnetic field $\bold{B}$                                 
\begin{equation}            
\bold{A}_{\rm C}(\bold{r},t)=\frac{\bold{\nabla}}{4\pi}\bold{\times}\int{\rm d}^3r'
\frac{\bold{B}(\bold{r}',t)}{|\bold{r}-\bold{r}'|}.
\label{AC}
\end{equation}
Stewart starts with the expression  (\ref{AC}) as an {\it ansatz} `suggested' by
the Helmholtz theorem, and then proceeds to prove it by substituting in (\ref{AC})
Jefimenko's expression for the magnetic field
in terms of the retarded current density and its partial time derivative
\cite {OJ} and obtaining, after some non-trivial algebra, an expression
for $\bold{A}_{\rm C}$ in terms of the current density
derived recently by Jackson \cite{Jack}.
In this comment, we give a more simple proof of the formula (\ref{AC})
using only the Helmholtz theorem.

According to the Helmholtz theorem \cite{Arf}, an arbitrary-gauge 
vector potential $\bold{A}$, as any three-dimensional vector 
field whose divergence and curl vanish at infinity, 
can be decomposed uniquely 
into a longitudinal part $\bold{A}_{\parallel}$,
whose curl vanishes, and  a transverse part  $\bold{A}_{\perp}$,
whose divergence vanishes
\begin{equation}
\bold{A}(\bold{r},t)=\bold{A}_{\parallel}(\bold{r},t)
+\bold{A}_{\perp}(\bold{r},t)\;\;\;\;\;\;
\bold{\nabla\times}\bold{A}_{\parallel}(\bold{r},t)=0\;\;\;\;\;\;
\bold{\nabla\cdot}\bold{A}_{\perp}(\bold{r},t)=0.
\label{HH}                  
\end{equation}
The longitudinal and transverse parts in (\ref{HH}) are given explicitly by
\begin{equation}
\bold{A}_{\parallel}(\bold{r},t)
=-\frac{\bold{\nabla}}{4\pi}\int{\rm d}^3r'\frac{\bold{\nabla}'
\bold{\cdot A}(\bold{r}',t)}{|\bold{r}-\bold{r}'|}\;\;\;\;\;\;
\bold{A}_{\perp}(\bold{r},t)
=\frac{\bold{\nabla}}{4\pi}\bold{\times}\int{\rm d}^3r'\frac{\bold{\nabla}'
\bold{\times A}(\bold{r}',t)}{|\bold{r}-\bold{r}'|}.
\label{longperp}
\end{equation}

Let us now decompose the vector potential
$\bold{A}$ in terms of the Coulomb-gauge vector potential
$\bold{A}_{\rm C}$ as follows:
\begin{equation}
\bold{A}(\bold{r},t)=[\bold{A}(\bold{r},t)-\bold{A}_{\rm C}(\bold{r},t)]
+\bold{A}_{\rm C}(\bold{r},t).
\end{equation}
If the curl of $[\bold{A}-\bold{A}_{\rm C}]$ vanishes, then, according to
equation (\ref{HH}) and the fact that the Coulomb-gauge
vector potential is by definition divergenceless,
the Coulomb-gauge vector potential $\bold{A}_{\rm C}$ is the transverse
part $\bold{A}_{\perp}$ of the vector potential $\bold{A}$.
But because the two vector potentials must yield the same
magnetic field, the curl of $[\bold{A}-\bold{A}_{\rm C}]$ does vanish
\begin{equation}
\bold{\nabla\times}[\bold{A}(\bold{r},t)-\bold{A}_{\rm C}(\bold{r},t)]
=\bold{\nabla\times}\bold{A}(\bold{r},t)            
-\bold{\nabla\times}\bold{A}_{\rm C}(\bold{r},t)
=\bold{B}(\bold{r},t)-\bold{B}(\bold{r},t)=0.
\end{equation}
Thus the Coulomb-gauge vector potential  is indeed
the transverse part of the vector potential $\bold{A}$ of any gauge.
Therefore, it can be expressed according to the second part of
(\ref{longperp}) and the fact that $\bold{\nabla\times A}=\bold{B}$ as
\begin{equation}
\bold{A}_{\rm C}(\bold{r},t)=\bold{A}_{\perp}(\bold{r},t)
=\frac{\bold{\nabla}}{4\pi}\bold{\times}\int{\rm d}^3r'\frac{\bold{\nabla}'
\bold{\times A}(\bold{r}',t)}{|\bold{r}-\bold{r}'|}
=\frac{\bold{\nabla}}{4\pi}\bold{\times}\int{\rm d}^3r'
\frac{\bold{B}(\bold{r}',t)}{|\bold{r}-\bold{r}'|}.
\label{AC2}
\end{equation}
The right-hand side of (\ref{AC2}) is expression (\ref{AC})
derived by Stewart.
                               
In closing, we note that there is an expression for the Coulomb-gauge 
scalar potential $V_{\rm C}$
in terms of the `instantaneous' electric field $\bold{E}$  that
is analogous to
expression (\ref{AC2}) for the Coulomb-gauge vector potential
\begin{equation}
V_{\rm C}(\bold{r},t)=\frac{1}{4\pi}\int{\rm d}^3r'\frac{\bold{\nabla}'
\bold{\cdot E}(\bold{r}',t)}{|\bold{r}-\bold{r}'|}.
\label{VC}
\end{equation}
This follows directly from the definition
$V_{\rm C}(\bold{r},t)=\int{\rm d}^3r'\rho(\bold{r}',t)/|\bold{r}-\bold{r}'|$
of the Coulomb-gauge scalar potential and the Maxwell equation
$\bold{\nabla\cdot E}=4\pi\rho$.
Expressions (\ref{AC2}) and (\ref{VC}) may be regarded 
as a `totally instantaneous gauge', but it would seem more appropriate to view
them as the solution to a problem that is inverse to that of calculating the
electric and magnetic fields
from given Coulomb-gauge potentials $\bold{A}_{\rm C}$ and $V_{\rm C}$
according to
\begin{equation}
\bold{E}=-\bold{\nabla}V_{\rm C}-\frac{\partial\bold{A}_{\rm C}}{c\partial t}
\;\;\;\;\;\;\bold{B}=\bold{\nabla\times A}_{\rm C}.
\label{EB}
\end{equation}
The first equation of (\ref{EB}) gives
directly the longitudinal part $\bold{E}_{\parallel}$  and transverse part
$\bold{E}_{\perp}$ of an electric field $\bold{E}$
in terms of the Coulomb-gauge potentials $V_{\rm C}$ and $\bold{A}_{\rm C}$ 
as $\bold{E}_{\parallel}=-\bold{\nabla}V_{\rm C}$ and 
$\bold{E}_{\perp}=-\partial\bold{A}_{\rm C}/c\partial t$ (the apparent paradox that 
the longitudinal part $\bold{E}_{\parallel}$ of a {\it retarded} electric field
$\bold{E}$ is thus a truly instantaneous field has been discussed recently in
\cite{Rohr}).

\end{document}